\documentclass[aps,prb,twocolumn,groupedaddress,showpacs]{revtex4-1}
\usepackage{graphicx}
\newcommand{\vsig}{\mbox{\boldmath$\sigma$}}
\newcommand{\vk}{{\bf k}}
\newcommand{\vg}{\mbox{\boldmath$g$}}
\newcommand{\vsk}{{\scriptstyle \mbox{\boldmath$k$}}}

\begin{document}

\preprint{}

\title{ Role of spin-orbit coupling on the electronic structure and properties of SrPtAs}
\author{S. J. Youn$^{1,2}$, S. H. Rhim$^2$, D. F. Agterberg$^3$, M. Weinert$^3$, A. J. Freeman$^2$}
\affiliation{$^1$ Department of Physics Education and Research Institute of Natural Science, Gyeongsang National University, Jinju 660-701, Korea
}
\affiliation{$^2$ Department of Physics and Astronomy, Northwestern University, Evanston, Illinois, 60208-3112, USA}
\affiliation{$^3$ Department of Physics, University of
Wisconsin-Milwaukee, Milwaukee, WI 53201-0413, USA}

\date{\today}

\begin{abstract}
The effect of spin-orbit coupling on the electronic structure of the
layered iron-free pnictide superconductor, SrPtAs, has been studied
using the full potential linearized augmented plane wave method.  The
anisotropy in Fermi velocity, conductivity and plasma frequency stemming
from the layered structure are found to be enhanced by spin-orbit
coupling.  The relationship between spin-orbit interaction and the lack
of two-dimensional inversion in the PtAs layers is analyzed within a
tight-binding Hamiltonian based on the first-principles calculations.
Finally, the band structure suggests that electron doping could increase $T_c$.
\end{abstract}
\maketitle

\section{Introduction}
\label{sec:intro}
The recent discovery of superconductivity in pnictides has attracted
extensive attention owing to their surprisingly high
$T_c$.\cite{jacs:kamihara08} While the highest $T_c$ so far is 56 K for
GdFeAsO,\cite{epl:wang08} a consensus on the pairing mechanism has not yet been
reached.\cite{physicaC:mazin09} This class of materials share a common
crystal structure, that is, the Fe square lattice.  While most of the
superconducting pnictides are Fe-based, pnictides without iron also
exhibit superconductivity, although $T_c$ is drastically lower than with
iron.  Recently, another superconducting pnictide, SrPtAs, has
been discovered, which is the first non-Fe based superconductor with a
hexagonal lattice rather than square lattice.\cite{nishikubo} Although
$T_c$=2.4K is lower than those of Fe-based pnictides, it
possesses interesting physics associated with its hexagonal crystal
structure.

\begin{figure}[b]
\includegraphics[width=\hsize]{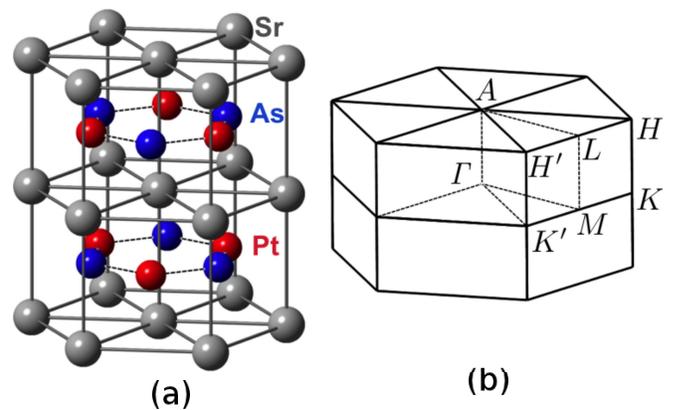}
\caption{\label{fig_crystal}
  (a) (Color online) Crystal structure of SrPtAs,
  where red, blue, and grey spheres denote Pt, As, and Sr atoms, respectively.
  (b) Brillouin zone of SrPtAs and high symmetry {\em k} points.
}
\end{figure}

SrPtAs crystallizes in a hexagonal lattice of ZrBeSi type with space
group $P6_3/mmc$ (No.194, $D_{6h}^4$) --- the same (non-symmorphic)
space group as the hcp structure --- with two formula units per
primitive cell.  As depicted in Fig.~\ref{fig_crystal}(a), its structure
resembles\cite{hoffmann} that of MgB$_2$ (with the symmorphic space
group $P6/mmm$, No.191, $D_{6h}^1$), with a double unit cell along the
{\em c} axis: the boron layers of MgB$_2$ are replaced by PtAs layers,
rotated by 60$^\circ$ in successive layers (responsible for the
non-symmorphic nature) and Mg is replaced by Sr. Although the crystal as
a whole has inversion symmetry (as do the Sr atoms), the individual PtAs
layers lack two-dimensional inversion.

Thus, SrPtAs differs from MgB$_2$ in two significant ways: ($i$) it
exhibits strong spin-orbit coupling (SOC) at the Pt ions and ($ii$) the
PtAs layers individually break inversion symmetry,  exhibiting only
$C_{3v}$ (or $D_{3h}$ if $z$-reflection is included) symmetry.  These
two properties play an important role in determining the band structure
and also affect the superconducting state. Assuming that the
superconductivity is largely determined by the two-dimensional PtAs
layers, the lack of inversion symmetry in the individual layers (which
we call ``broken local inversion symmetry'') opens up the possibility to
see the unusual physics associated with non-centrosymmetric
superconductors.\cite{frigeri}
With large spin-orbit
coupling, nominally $s$-wave non-centrosymmetric superconductors exhibit
spin-singlet and spin-triplet mixing,\cite{gor01,yua06} enhanced spin
susceptibilities,\cite{gor01,fri04} enhanced Pauli limiting
fields,\cite{frigeri} non-trivial magnetoelectric effects and
Fulde-Ferrell-Larkin-Ovchinnikov(FFLO)-like states in magnetic
fields,\cite{bar02,agt03,kau05,sam08,ede96,yip02,fuj05} and Majorana
modes.\cite{sat09} The local inversion symmetry breaking, together with
a SOC that is comparable to the $c$-axis coupling, suggests that
SrPtAs will provide an ideal model system to explore related effects in
centrosymmetric superconductors.\cite{youn2011,prb11:mh_fisher}


In this paper we discuss the electronic structure of SrPtAs, including
spin-orbit coupling, which was neglected in a previous theoretical
study.\cite{shein} In Sec.~\ref{sec:details}, we describe details of the
calculations.  The effects of SOC on the bands, the Fermi surface,
density of states, and transport properties at the Fermi surface of
SrPtAs are presented in Sec.~\ref{sec:results}, along with a
tight-binding analysis.  Finally, we suggest an enhanced $T_c$ might be
possible via doping.

\section{Method}
\label{sec:details}
First-principles calculations are performed using the full-potential
linearized augmented plane wave (FLAPW) method\cite{wimmer,weinert} and
the local density approximation (LDA) for the exchange-correlation
functional of Hedin and Lundqvist.\cite{hedin} Then, SOC is included by
a second variational method.\cite{macdonald} Experimental lattice
constants, $a$= 4.244\AA\ and $c$= 8.989\AA, are used.\cite{wenski}
Cutoffs used for wave function and potential representations are 196 eV
and 1360 eV, respectively.  Muffin-tin radii are 2.6, 2.4, and 2.1 a.u
for Sr, Pt, and As, respectively.  Semicore electrons such Sr $4p$ and
As $3d$ are treated as valence electrons, which are explicitly
orthogonalized to the core states.\cite{explicit} Brillouin zone
summations were done with 90 {\em k} points in the Monkhorst-Pack
scheme,\cite{kpts} while the density of states are obtained by the
tetrahedron method.\cite{blochl} Although most of the calculations were
performed using LDA, some results were also done with the GGA as
well.\cite{gga} In order to calculate the Fermi velocity and plasma
frequency for in-plane and out-of-plane contributions, eigenvalues from
self-consistent calculations are fitted by a spline method over the
whole Brillouin zone.\cite{koelling86,spline,boltztrap}

\section{Results}
\label{sec:results}
\begin{figure}[b]
\includegraphics[width=\hsize]{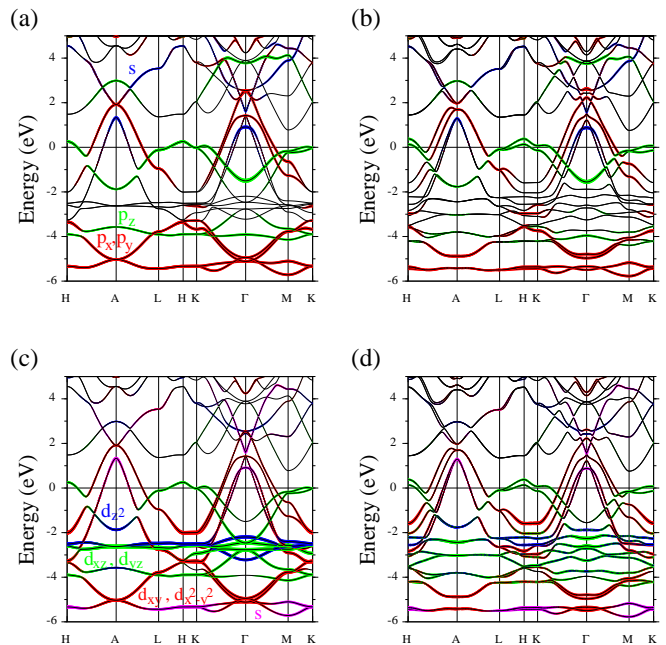}
\caption{\label{fig:band}
(Color online) Band structure of SrPtAs (a),(c) without and (b),(d) with
spin-orbit coupling.  In (a) and (b), As $p_{x,y}$ and $p_z$ orbitals
are shown in red and green, respectively, and As $s$ in blue.  In (c) and
(d), Pt $(d_{xy},d_{x^2-y^2})$, $(d_{xz},d_{yz})$, and $d_{z^2}$
orbitals are presented in red, green, and blue, respectively.
Contribution from Pt {\em s} is shown in purple.
}
\end{figure}

The band structure of SrPtAs is presented in Fig.~\ref{fig:band} along
the symmetry lines shown in Fig.~\ref{fig_crystal}(b).
Plots in the left (right)
column are without (with) spin-orbit coupling, and plots in the upper
and lower rows are the same but highlighted for As {\em p} and Pt {\em
d} orbitals, respectively.  Our energy bands and Fermi surfaces without
SOC agree with those of Ref.~[\onlinecite{shein}].  The main band of Sr
5$s$ origin is located far above the Fermi level ($E_F$),
consistent with Zintl's scheme\cite{schafer} that Sr donates electrons
to the PtAs layer and behaves almost like an inert Sr$^{2+}$ ion.  {\em
Without} SOC, bands on the zone boundary face, the $k_z=\pi/c$
plane($H-A-L-H$), exhibit four fold degeneracy --- two from spin and
the other two from two different PtAs layers --- as a consequence of the
non-symmorphic translations along the $c$-axis, just as for the hcp
structure.  For the $k_z=0$ plane ($K-\Gamma-M-K$), there is no such
symmetry-dictated degeneracy due to the two equivalent layers, but
instead the magnitude of splitting is proportional to the inter-layer
coupling.  With SOC, the bands change markedly: The four-fold degeneracy
on the zone boundary face is reduced to a two fold pseudospin degeneracy
due to inversion symmetry, whereas bands along the $A-L$ line keeps the
fourfold degeneracy as a consequence of time-reversal
symmetry.\cite{mattheiss}

In a simple atomic picture, the SOC Hamiltonian is $H_{soc} = \delta
{\bf L}\cdot\vsig $, where $\delta$ represents the strength of the SOC.
Values of the SOC strength derived from the calculations are
$\delta_{Pt}$=0.32 eV for the Pt {\em d} orbitals and $\delta_{As}$=0.23
eV for As {\em p} orbitals, which are used in later discussions.  The
SOC splitting at the $A$ point for Pt $d_{xz},d_{yz}$
($d_{xy},d_{x^2-y^2}$) orbitals is 0.59 (0.54) eV, whereas for As
$p_{x,y}$ the splitting 0.28 eV.

To gain insight into the effect of spin-orbit coupling, we consider a
simple tight-binding theory for a single PtAs layer.  A single PtAs layer lacks a center of inversion symmetry and therefore allows an anti-symmetric spin-orbit coupling of the form
\begin{equation}
  \mathcal{H}_{so} = \sum_{\vsk,s,s'}
  \vg_{\vsk} \cdot \vsig_{ss'} c_{\vsk s}^{\dag} c^{\phantom{\dag}}_{\vsk s'}
\label{eq-1}
\end{equation}
exists, where $ c_{\vsk s}^{\dag} $ ($ c^{\phantom{\dag}}_{\vsk s} $) creates (annihilates)
an electron with  momentum $ \vk $  and pseudo-spin $s$, and $\vsig$ denote the Pauli matrices. Time-reversal symmetry imposes $\vg_{\vsk}=-\vg_{-\vsk}$. Here we find the form of $\vg_{\vsk}$ through a consideration of the coupling between the Pt $d_{x^2-y^2},d_{xy}$ orbitals
and the As $p_x,p_y$ orbitals. These orbitals give rise to the Fermi surfaces with
cylindrical topology around the $\Gamma$--$A$ line. On the Pt sites, we
define states $|d\pm,s>=|d_{x^2-y^2},s>\pm i |d_{xy},s>$ and on the As
sites we define the states $|p\pm,s>=|p_x,s>\pm i |p_y,s>$ ($s$ denotes
spin). On a single Pt or As site, the spin-orbit coupling ${\bf L}\cdot
{\bf S}$ has only $L_z S_z$ with non-zero matrix elements in these
subspaces. This splits the local four-fold degeneracy into two pairs so
that for Pt (As) $|d+,\uparrow >$ and $|d-,\downarrow >$ ($|p+,\uparrow
>$ and $|p-,\downarrow >$) form one degenerate pair while
$|d+,\downarrow >$ and $|d-,\uparrow >$ ($|p-,\uparrow >$ and
$|p+,\downarrow >$) form the other degenerate pair. For both pairs,
time-reversal symmetry is responsible for the degeneracy and we can
label the two degenerate partners of each pair through a pseudo-spin
index. We include a spin-independent nearest neighbor hopping between
the As and Pt sites. This yields the following tight binding Hamiltonian
in {\em k}-space
\begin{equation}
\mathcal {H}_0 = \sum_{\vsk,s} \Psi_s^{\dagger}(\vsk) H_s(\vk) \Psi_s({\vsk}),
\label{eq-2}
\end{equation}
where $\Psi_s({\vsk})=(c_{\vsk,d+,s},c_{\vsk,d-,s},c_{\vsk,p+,s},c_{\vsk,p-,s})^T$, $s$ is the spin label ($s=\{\uparrow,\downarrow\}$). For spin up, $H_{\uparrow}(\vk)$ is (to get $H_{\downarrow}(\vk)$ change the sign of $\delta_{Pt}$ and $\delta_{As}$)
\begin{widetext}
\begin{equation}
\left(
    \begin{array}{cccc}
      \epsilon_d+\delta_{Pt} & 0 & t(1+\nu^*e^{-i\vk\cdot{\bf T}_3}+\nu e^{i\vk\cdot{\bf T}_2})& \tilde{t}(1+e^{-i\vk\cdot{\bf T}_3}+e^{i\vk\cdot{\bf T}_2}) \\
      0& \epsilon_d-\delta_{Pt}&  \tilde{t}(1+e^{-i\vk\cdot{\bf T}_3}+e^{i\vk\cdot{\bf T}_2})& t(1+\nu e^{-i\vk\cdot{\bf T}_3}+\nu^* e^{i\vk\cdot{\bf T}_2}) \\
      t(1+\nu e^{i\vk\cdot{\bf T}_3}+\nu^* e^{-i\vk\cdot{\bf T}_2}) &  \tilde{t}(1+e^{i\vk\cdot{\bf T}_3}+e^{-i\vk\cdot{\bf T}_2})& \epsilon_p+\delta_{As}&0  \\
      \tilde{t}(1+e^{i\vk\cdot{\bf T}_3}+e^{-i\vk\cdot{\bf T}_2})&t(1+\nu^* e^{i\vk\cdot{\bf T}_3}+\nu e^{-i\vk\cdot{\bf T}_2})&0&\epsilon_p-\delta_{As} 
    \end{array}
  \right)
\end{equation}
\end{widetext}
where $\nu=e^{i2\pi/3}$, ${\bf T}_1=a(1,0)$, ${\bf T}_2=a(-1/2,\sqrt{3}/2)$, ${\bf T}_3=a(-1/2,-\sqrt{3}/2)$,
$t$  and $\tilde{t}$ are the hopping parameters between the As $p$ and Pt $d$ orbitals, and
$\delta_{Pt}$ ($\delta_{As}$) is the atomic SOC parameter for Pt $d$(As $p$) orbitals.
To find an effective Hamiltonian for the Pt $d$-orbitals, we assume that $|\epsilon_d-\epsilon_p|$ is the largest energy scale and treat all other parameters ($t,\tilde{t},\delta_{Pt},\delta_{As}$) as perturbations. This yields the following effective Hamiltonian \begin{equation}
\mathcal {H}_{Pt} = \sum_{\vsk,s} \Psi_{Pt,s}^{\dagger}(\vsk) H_{Pt,s}(\vk) \Psi_{Pt,s}({\vsk}),
\label{eq-3}
\end{equation}
where $\Psi_{Pt,s}({\vsk})=(c_{\vsk,d+,s},c_{\vsk,d-,s})^T$, $s$ is the spin label, and
\begin{widetext}
\begin{equation}
H_{Pt,\uparrow}=\left(
    \begin{array}{cc}
     \epsilon_d+\delta_{Pt} +t_1\sum_{i}\cos(\vk \cdot {\bf T}_i)+t_2\sum_{i}\sin(\vk \cdot {\bf T}_i)& t_3[\cos(\vk \cdot {\bf T}_1)+\nu \cos(\vk \cdot {\bf T}_2)+\nu^*\cos(\vk \cdot {\bf T}_3)] \\
     t_3[\cos(\vk \cdot {\bf T}_1)+\nu^*\cos(\vk \cdot {\bf T}_2)+\nu\cos(\vk \cdot {\bf T}_3)] &  \epsilon_d-\delta_{Pt}+t_1\sum_{i}\cos(\vk \cdot {\bf T}_i)-t_2\sum_{i}\sin(\vk \cdot {\bf T}_i)
    \end{array}
  \right)
\end{equation}
\end{widetext}
where $t_1=2[\tilde{t}^2+\cos(2\pi/3)t^2]/(\epsilon_d-\epsilon_p)$, $t_2=-2\sin(2\pi/3)t^2/(\epsilon_d-\epsilon_p)$, and $t_3=-2t\tilde{t}/(\epsilon_d-\epsilon_p)$, and the sum over $i$ is over the three vectors ${\bf T}_1, {\bf T}_2$, and ${\bf T}_3$. The Hamiltonian $H_{Pt,\uparrow}$ yields the dispersions $\epsilon(\vk)=t_1(\vk)\pm\sqrt{(\delta_{Pt}+t_2(\vk))^2+|t_3(\vk)|^2}$ where $t_1(\vk)=t_1\sum_i\cos(\vk \cdot {\bf T}_i)$, $t_2(\vk)=t_2\sum_i\sin(\vk \cdot {\bf T}_i)$, and $t_3(\vk)=t_3[\cos(\vk \cdot {\bf T}_1)+\nu \cos(\vk \cdot {\bf T}_2)+\nu^*\cos(\vk \cdot {\bf T}_3)] $. The SOC contribution that lifts the four-fold degeneracy into two
bands $E_{\vk,\pm}=E_\vk\pm |g(\vk)|$ at $k_z=\pi/c$ can be found to order $\delta_{Pt}/\sqrt{\delta_{Pt}^2+|t_2(\vk)|^2+|t_3(\vk)|^2}$ and is given by

\begin{eqnarray}
\vg(\vk)=&&\hat{z} \frac{t_2\delta_{Pt}}{\sqrt{\delta_{Pt}^2+|t_2(\vk)|^2+|t_3(\vk)|^2}}\nonumber\\
         &&\times[\sin(\vk \cdot {\bf T}_1)+\sin(\vk \cdot {\bf T}_2)+\sin(\vk \cdot {\bf T}_3)]
\label{eq_4}
\end{eqnarray}
where we have also included the contribution from
$H_{Pt,\downarrow}(\vk)$. Note that this $\vg$ leads to pseudo-spin interaction in Eq.~\ref{eq-1} denoted by
$\sigma_z$. We emphasize that this $\sigma_z$ operates on
pseudo-spin, not actual spin, the up and down pseudo-spin states are
related by time-reversal symmetry (for example, the local states
$|d+,\uparrow >$ and $|d-,\downarrow >$ form a pseudo-spin pair). The expression for $\vg$
clearly reveals how the interplay between the atomic SOC
($\delta_{Pt}$) and the broken inversion symmetry
($\vg(\vk)=-\vg(-\vk)$) of a single PtAs layer leads to the relevant
band SOC. Note that this single-layer band SOC will be of opposite signs
for the two inequivalent PtAs layers in the unit cell. Further,
the band spin-orbit splitting will have additional contributions from
terms of order
$\delta_{Pt}/(\epsilon_d-\epsilon_p)$ and
$\delta_{As}/(\epsilon_d-\epsilon_p)$ that were neglected in the above
derivation. However,  these additional contributions do not
qualitatively change the results. Similar considerations apply for the
other Pt $d$-orbitals.

\begin{figure}[b]
\includegraphics[width=\hsize]{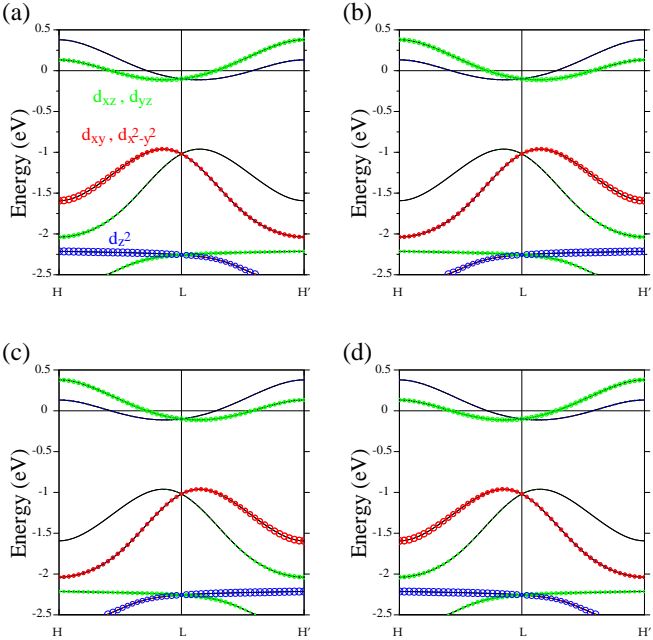}
\caption{\label{fig_rashba}
(Color online) SOC splitting of bands.
  (a) and (b) [(c) and (d)] show components from Pt 5$d$ orbitals for
upper [lower] Pt atoms. Left column[(a) and (c)] is for spin-up
components and right[(b) and (d)] is for spin-down components. Radius of
circles is proportional to the magnitude of the components.}
\end{figure}

The previous paragraph considered a single PtAs layer. To complete the description, the coupling between the two inequivalent layers must be included. For the Pt $d_{x^2-y^2},d_{xy}$ orbitals considered above, the nearest neighbor inter-layer hopping matrix between $|d+,s >$ states is  (the same expression appears for $|d-,s >$ states)
\begin{equation}
\epsilon_c({\bf k})= t_c \cos(k_zc/2)(1+e^{-i{\bf k}\cdot {\bf T}_3}+e^{i{\bf k}\cdot {\bf T}_2}).
\end{equation}
Including this inter-layer hopping leads to the following Hamiltonian
\begin{eqnarray}
  \mathcal {H}_{\pm,s} = \sum_{\vsk} \Psi^{\dagger}_{\pm,s}(\vsk)\Big\{ [\epsilon_{\pm}({\vsk})-\mu] \sigma_0\tau_0 + \vg(\vk)\cdot \vsig \tau_z \nonumber\\
+Re[\epsilon_c(\vsk)]\sigma_0\tau_x+ Im[\epsilon_c(\vsk)]\sigma_0\tau_y\Big\}\Psi_{\pm}({\vsk,s'}),
\label{eq-2}
\end{eqnarray}
where $\Psi_{\pm,s}({\vsk})=(c_{\pm \vsk \uparrow 1,s},c_{\pm \vsk \downarrow 1,s},c_{\pm \vsk \uparrow 2,s},c_{\pm \vsk \downarrow 2,s})^T$, 1,2 denote the two inequivalent PtAs layers,  $\sigma_i$ ($\tau_i$) are Pauli matrices that operate on the pseudo-spin (layer) space, $\epsilon_{\pm}=t_1(\vk)\pm\sqrt{(t_2(\vk))^2+|t_3(\vk)|^2}$, and $\vg(\vk)$ is given in Eq.~(\ref{eq_4}) (the $\tau_z$ matrix describes the sign change of $\vg$ on the two layers).
This Hamiltonian can be diagonalized with resulting dispersion relations $\epsilon({\bf k})=\epsilon_{\pm}({\bf k})\pm\sqrt{|\epsilon_c({\bf k})|^2+\vg^2(\vk)}$ and each state is 2-fold degenerate due to time-reversal symmetry (Kramers degeneracy). Note that the tight binding theory described above suggests that eigenstates of $S_z$ are also eigenstates of the single electron Hamiltonian. However, inter-layer coupling terms can lead to additional terms that do not commute  with $S_z$. The band structure suggests that these terms are not large for the states near the Fermi surface.

The SOC found in in Eq.~(\ref{eq_4}) has
opposite sign for the different layers as well as for the pseudo-spin
direction.  This is demonstrated in Fig.~\ref{fig_rashba}, where the
band structure is resolved by layer and spin for $H$--$L$--$H'$, where
$L$ is one of the time-reversal invariant momentum (TRIM) points, and
the lines $H$--$L$ and $L$--$H'$ are related both by a mirror plane and,
and more relevant to the discussion, by a combination of inversion and
reciprocal lattice vectors.  In Fig. \ref{fig_rashba}, spin up[(a)] and
down[(b)] components of the Pt {\em d} orbital from the upper PtAs layer
are marked in different colors, while components from the other PtAs
layers are shown in (c) and (d). As expected from the tight-binding
analysis, Figs.~\ref{fig_rashba}(a),(d) [and similarly for (b) and (c)]
appear the same since they correspond to spatial inversion and
time-reversal (opposite spin).  Despite the presence of a global
inversion
center, the locally broken inversion symmetry in PtAs is evident
in Fig.~\ref{fig_rashba} where the spin degeneracy is broken in a single
layer, for which the consequences are nontrivial.\cite{youn2011} (This
result is not surprising or unexpected since in the limit that the
coupling between PtAs layers vanishes, i.e., the separation goes to
infinity, the single layer result must be recovered.)
This spin separation, however, does not result in magnetism because of spin
compensation in each layer.   These antisymmetric splittings
($\vg(\vk)=-\vg(-\vk)$) occur not only at the $L$ point but also at
the other TRIM points\cite{fu} --- $\Gamma, A$, $M$,
and the $A$-$L$ line --- in the hcp structure.

\begin{figure}[b]
\includegraphics[width=\hsize]{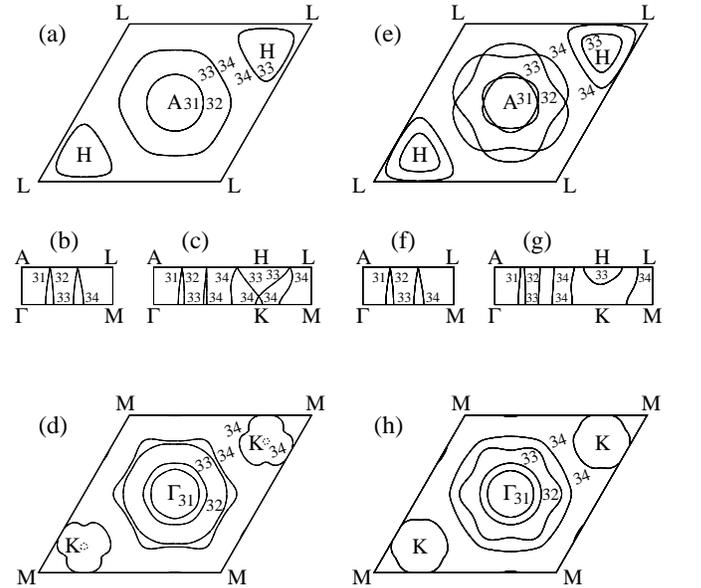}
\caption{\label{fig:fermi}
Cross section of the Fermi surface (a)-(d):  without and (e)-(h): with
spin-orbit coupling. Cross section along the zone boundary face,
$k_z=\pi/c$ [(a),(e)] and at the zone center, $k_z=0$ [(d),(h)]. (b),(c)
are contours along the vertical plane shown with their corner points in
the Brillouin zone, where (f),(g) are those with SOC included.  Numbers
next to Fermi surface sheets indicate band indices.
%
}
\end{figure}

As expected from the layer structure, anisotropy occurs in the Fermi
surfaces, conductivity, and plasma frequency.  Two-dimensional cross
sections of the Fermi surfaces are shown in Fig.~\ref{fig:fermi}, both
without [(a)-(d)] and with [(e)-(h)] SOC.  Contours in the $k_z=\pi/c$
[(a),(e)] and $k_z=0$ [(d),(h)] planes clearly exhibit the consequence
of the crystal symmetry: hexagonal symmetry around $\Gamma$ and $A$ and
trigonal symmetry around $K$ and $H$.  All sheets exhibit almost
two-dimensional cylindrical features except for a small pocket around
$H$.  This anisotropy is further exemplified by transport properties
which will be discussed later.  Moreover, because of the
symmetry-dictated degeneracy due to the non-symmorphic symmetry, there
is no spin-orbit splitting along the time-reversal invariant direction,
$A$-$L$, as seen in Fig.~\ref{fig:fermi}(e).  Comparing
Figs.~\ref{fig:fermi}(c) and (g), the SOC appears to make the Fermi
surfaces more cylindrical, consequently enhancing the two dimensional
character of the Fermi surfaces.

All the Fermi surfaces are hole-like after turning on SOC, in sharp
contrast to other pnictide superconductors with two electron-like Fermi
surfaces and two hole-like Fermi surfaces.\cite{sing2008}
Instead of electron-like and hole-like Fermi surfaces, they are
distinguished by orbital character.  Sheets around the $\Gamma$-$A$ line
consist of $\sigma$ orbitals of the PtAs layer, As $p_{x,y}$ and
Pt $d_{xy,x^2-y^2}$, while sheets around the $K$-$H$ line are from $\pi$ orbitls, As $p_z$
and Pt $d_{xz,yz}$.  Two kinds of Fermi surfaces with different orbital
character might give rise to a two energy gap superconductor in
SrPtAs.

The anisotropy due to the layered structure is further manifested in the
average Fermi velocities and plasma frequencies.  Neglecting SOC, the
in-plane and out-of-plane Fermi velocities are $\langle
v_{x,y}^2\rangle^{1/2}$=3.72$\times 10^7$ cm/s and $\langle
v_z^2\rangle^{1/2}$=1.02$\times 10^7$ cm/s, respectively, and the plasma
frequencies are $\Omega_{x,y}$=5.70 eV and $\Omega_z$=1.57 eV.  The
anisotropy ratio, defined as a ratio of conductivities between
in-plane and out-of-plane components, is 13.3, assuming an isotropic
scattering rate. With SOC, the anisotropy is enhanced: $\langle
v_{x,y}^2\rangle^{1/2}$=3.76$\times 10^7$ cm/s and $\langle
v_z^2\rangle^{1/2}$=6.78$\times 10^6$ cm/s; $\Omega_{x,y}$=5.57 eV,
$\Omega_z$=1.00 eV, and the anisotropy ratio increases to a much higher
value of 30.8.
The decrease of $v_z$ by 33\% by SOC is consistent with the enhanced two
dimensional character of the Fermi surfaces. Table~\ref{table1}
summarizes contributions from each Fermi surface.  The largest
contribution to the density of states at the Fermi level, $N(0)$, comes
from the 34th band at around $K$, which is due to the low velocity at
the Fermi surface. The anisotropy ratio is usually much larger than 1
except for the small hole pocket around $H$.

\begin{table}[b]
\caption{\label{table1}
Fermi surface properties with SOC included for each surface: Density of
states, $N(0)$ (states/eV/spin); average velocities, $\langle
v_x^2\rangle^{1/2}$, $\langle v_z^2\rangle^{1/2}$ ($10^7$cm/s);
anisotropy ratio, $\langle v_x^2\rangle/\langle v_z^2\rangle$; and
plasma frequencies, $\Omega_x$, $\Omega_z$  (eV).  The numbers 31,32,33,
and 34 represent band indices and $\Gamma$ and K in the parenthesis
represent the locations of the Fermi surface.
}
\begin{ruledtabular}
\begin{tabular}{cccccccc}
     &   31     &     32 & 33($\Gamma$)  &      34($\Gamma$)  &      33(K)  &      34(K) &   Total \\
\hline
    $N(0)$  & 0.085     &     0.107     &     0.209     &     0.346     &     0.267     &     0.943  &     1.898\\
$\langle v_x^2\rangle^{1/2}$    &      7.03 & 6.79 & 5.92 & 4.35 & 1.25 & 1.68 &  3.76 \\
$\langle v_z^2\rangle^{1/2}$   &    1.03 & 0.67 & 0.47 & 1.04 &  1.37 & 0.41 & 0.678 \\
 $\langle v_x^2\rangle/\langle v_z^2\rangle$  &  46.9    & 103  &  159 &  17.3 &  0.83 &  16.9  &  30.8 \\
$\Omega_x$     &     2.20  &     2.39 &      2.91   &    2.74  &     0.69 & 1.75 & 5.57 \\
$\Omega_z$     &     0.32 &     0.24 &     0.23 &     0.66 &    0.76 & 0.43 & 1.00 \\
\end{tabular}
\end{ruledtabular}
\end{table}

\begin{figure}[tb]
\centering
\includegraphics[width=1.5\hsize]{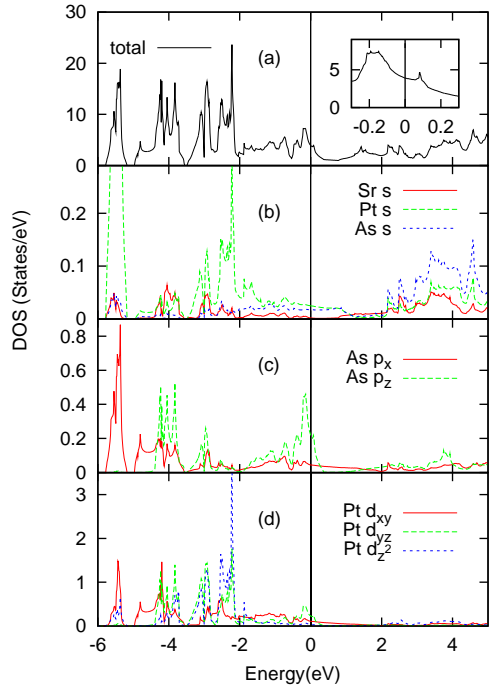}
\caption{\label{fig:dos}
(Color online) Density of states (DOS) of SrPtAs: (a) total DOS,
  (b) {\em s} orbitals, (c) As {\em p} orbitals, and (d) Pt {\em d} orbitals.
}
\end{figure}

The total density of states (DOS) and orbital decomposed partial DOS
are presented in Fig.~\ref{fig:dos}.
The states around $E_F$ arise mainly from As {\em p} and Pt {\em d}.
The nonbonding Pt $d_{z^2}$ bands are located at around -2.2 eV, and
because they
hybridize little with other orbitals, they give rise to a peak in the
DOS.  In contrast, $\pi$-bonding orbitals such as As $p_z$ and Pt
$d_{xz,yz}$ are located around the Fermi level with a wide band width.
In particular, the van Hove singularity (vHS) coming from a saddle point
near $K$ just above $E_F$ (inset of Fig.~\ref{fig:dos}(a))exhibits two
dimensional character since a 2D vHS gives rise to a singularity in the
DOS while a 3D vHS gives only a discontinuity in slope.  Raising $E_F$
to vHS, which might be realized by electron doping via Sr layers, could
potentially increase $T_c$: Assuming rigid bands, we estimate that
electron doping by roughly 25\% will lift $E_F$ to the vHS, enhancing
the DOS at the Fermi level by 6\%.  Further, using a weak-coupling form for $T_c$ ($T_c=1.14T_D e^{-1/N(0)V}$), assuming that the pairing
potential $V$ does not change by electron doping, and assuming Debye temperature of $T_D=
229 K$,\cite{private2} we estimate that $T_c$ could be enhanced to as high as 3.8K
by electron doping.

Applying hydrostatic pressure could also raise the DOS and could, as
often happens, enhance $T_c$.  Rh is a candidate to supply a chemical
pressure to SrPtAs since it has a similar electronegativity as Pt but a
smaller ionic radius.  However, the crystal structure is sensitive to
the constituent atoms; for example, SrPtSb where As is replaced by Sb
has an AlB$_2$-type structure, while YPtAs, where Sr is replaced by Y
with more electrons, has a hexagonal structure with four slightly
puckered PtAs layers in a unit cell.\cite{hoffmann}

Finally we consider the possibility of simple collinear magnetic
solutions. The antiferromagnetic (AFM) phase, where the moments in a
PtAs layer are aligned, by antiparallel to those in adjacent layers, is
favored by 0.23 meV (0.49 meV in GGA) per formula unit,
than the non-magnetic phase; the ferromagnetic orientation converged to
the non-magnetic solution.
Magnetic moments are given in Table~\ref{table2}.
While the energy differences and calculated moments are too small to
make definitive statements regarding magnetic phases in SrPtAs,
the material appears to be near a magnetic instability.

\begin{table}[b]
\caption{\label{table2}
Spin $\mu_s$ and orbital $\mu_l$ magnetic moments and total magnetic moment $\mu_t$ in unit of $\mu_B$.
}
\begin{ruledtabular}
\begin{tabular}{cccc}
& $\mu_s$ & $\mu_l$ & $\mu_t$ \\
\hline
Pt &0.014 &0.064 & 0.078\\
As &0.006& -0.027& -0.021\\
\end{tabular}
\end{ruledtabular}
\end{table}

\section{Summary}
\label{summary}
First-principles calculations of the electronic structure of SrPtAs have
been presented with SOC fully taken into account.  The role of SOC on
the electronic structure is manifested in the energy bands and Fermi
surfaces. The important physics originates from two factors: strong SOC
in Pt atoms and locally broken inversion symmetry in PtAs layers.
We have constructed a tight-binding Hamiltonian based on the
self-consistent electronic structure that provides insight into the SOC.
Sheets of the Fermi
surface are spatially well separated in the Brillouin zone: cylindrical
Fermi surfaces with $\sigma$-character at the zone center (around
$\Gamma$-$A$)
and two Fermi surfaces, {\it i.e.,} a pocket and a cylinder, with $\pi$-character at the zone
corner (around $K$-$H$).  All the Fermi surfaces are hole-like which
distinguishes this material from other pnictide superconductors.  The
transport properties are highly anisotropic between $x,y-$ and $z-$
directions.  Rh is suggested for a positive pressure effect to increase
$T_c$.  Furthermore, the van Hove singularity is shown in the DOS above
$E_F$.  Assuming rigidity of bands, we predict that $T_C$ increases up
3.4 K with 25\% doping, which may be achieved by chemical doping in
place of the Sr atom.

\begin{acknowledgments}
  SJY and SHR are indebted to Hosub Jin for fruitful discussions.
  SJY acknowledges the sabbatical research Grant by Gyeongsang National University.
  SHR and AJF are supported by the Department of Energy (DE-FG02-88ER45382).
  DFA is supported by NSF grant DMR-0906655. MW is supported by NSF
DMR-1105839.
\end{acknowledgments}
\bibliography{spa_ref3}


\end{document}